\documentclass[twocolumn,nofootinbib,showpacs,preprintnumbers,amsfonts]
{revtex4}

\usepackage{dcolumn}
\usepackage{hyperref}
\newcommand{\be}[3]{\begin{equation}  \label{#1#2#3}}
\newcommand{\bea}[3]{\begin{eqnarray}  \label{#1#2#3}}
\newcommand{\ee}{\end{equation}}
\newcommand{\eea}{\end{eqnarray}}
\newcommand{\ba}{\begin{array}}
\newcommand{\ea}{\end{array}}

\newcommand{\1}{{\mathbb I}}
\newcommand{\N}{${\cal N}\, $}

\newcommand{\haken}{\mathbin{\hbox to 8pt{%
                 \vrule height0.4pt width7pt depth0pt
                 \kern-.4pt
                 \vrule height4pt width0.4pt depth0pt\hss}}}

\begin{document}


\preprint{UPR-1070-T, AEI-2004-016, hep-th/0403049}

\title{General \N = 1 Supersymmetric Flux Vacua  of 
 (Massive) Type IIA String Theory  }
 
\author{Klaus Behrndt}
  \email{behrndt@aei.mpg.de}
 \altaffiliation[on leave from:]{\em Albert-Einstein-Institut, 
Am M\"uhlenberg 1, 14476 Golm,
Germany.}
\author{Mirjam Cveti{\v c}}%
 \email{cvetic@cvetic.hep.upenn.edu}
\affiliation{%
Department of Physics and Astronomy, \\
University of Pennsylvania, Philadelphia, PA 19104-6396, USA}

\date{\today}

\begin{abstract}
We derive conditions for the existence of four-dimensional \N=1
supersymmetric flux vacua of massive type IIA string theory with
general supergravity fluxes turned on.  For an SU(3) singlet Killing
spinor, we show that such flux vacua exist only when the internal
geometry is nearly-K\"ahler. The geometry is not warped, all the
allowed fluxes are proportional to the mass parameter and the dilaton
is fixed by a ratio of (quantized) fluxes.  The four-dimensional
cosmological constant, while negative, becomes small in the vacuum
with the weak string coupling.

\end{abstract}


\maketitle


Insights into four-dimensional \N = 1 supersymmetric vacua of M- and
string theory with non-Abelian gauge sectors and chiral matter are of
phenomenological interest: they may provide an important link between
the M-theory and particle physics, describing the Standard Model
and/or Grand Unified models.  In particular, intersecting D6-brane
constructions \cite{Blumetal,Ibanezetal} on Type IIA orientifolds
provide a intriguing avenue to construct semi-realistic particle
physics. [For a recent review see \cite{770} and references therein;
for first examples of \N=1 supersymmetric quasi-realistic models see
\cite{CSUI,CSUII}]

On the other hand more general compactifications of string theory in
the presence of supergravity fluxes can lift the continuous
moduli space of string vacua, while still preserving some
supersymmetries and hence might be an essential input in realistic
compactifications of string/M-theory.  These fluxes generate a back
reaction onto the geometry, which in the simplest case results in a
non-trivial warping, but in general the internal space ceases to be
Calabi-Yau.  Unfortunately, the explicit metric and fluxes are
known only for very few examples. 

One of the important phenomenological goals in the constructions of
general N=1 supersymmetric vacua is the implementation of
(supergravity) flux compactifications, which would yield the moduli
stabilization, along with the (probe) D-brane configurations, which
would yield the non-Abelian gauge group structure and chiral matter in
four-dimensions.  On one hand, a better understanding of the resulting
internal geometry can be achieved by relating the fluxes to specific
non-trivial torsion components, which can be classified with respect
to the structure group of the internal manifold, for a recent review
see \cite{650}.  The Killing spinor has to be a singlet under the
structure group which for a six-dimensional space is at most SU(3).
[However non-zero fluxes often require a reduction of the structure
group down to SU(2).]  It is this maximal SU(3) structure group that
plays an important role in the construction of D-brane configurations
with chiral matter.  In particular, within Type IIA string theory the
chiral supermultiplets appear at the intersections of two D6-branes,
whose three-cycles are related to each other by an SU(3) rotations
\cite{130} and supersymmetry is preserved if the Killing spinor is a
singlet under the SU(3) rotations.

The goal of this letter is to derive within massive type IIA string
theory explicit constraints for the most general four-dimensional N=1
supersymmetric flux compactification whose internal space maintains
SU(3) structure.  These results thus provide an important stepping
stone toward implementation of chiral theories from D-brane
configurations along with the moduli stabilization within Type IIA
theory.  In particular we show that such vacua exist for massive Type
IIA string theory only when the internal geometry is
nearly-K\"ahler. The geometry is not warped, the allowed flux
components are related to the mass parameter and the dilaton is fixed
by a ratio of supergravity fluxes.  The four-dimensional cosmological
constant is negative, however in the weak coupling limit it becomes
arbitrarily small. We also give explicit examples with nearly-K\"ahler
internal geometry.  In this letter we primarily summarize results,
technical details as well as the analysis of SU(2) structures shall be
presented in \cite{333}, see \cite{100}.


In massive IIA string theory \cite{110p}, the NS-NS 2-form and the RR 1-form
potential combine into a gauge invariant (massive) 2-form $F$ which
fixes also the Chern-Simons part of the RR 4-form $G$
\be282
F = m B + dA \ , \ G = dC + {3 \over m} F \wedge F
\ee
were $C$ is the RR 3-form potential. These forms are not closed but
\be732 
dF = m H \ , \ dG = 6 F \wedge H \ .
\ee 
In the massless case, $F$ and $G$ are independent fields, where $G$ is
still not closed but $F=F^{(0)}= dA$ is exact.  Unbroken supersymmetry
requires the existence of at least one Killing spinor $\epsilon$,
which is fixed by the vanishing of the fermionic supersymmetry
transformations. These variations have been first setup for massive
type IIA supergravity in Einstein frame \cite{110p}, but we will use
the string frame where the fermionic variations read
\be040
\ba{l}
\delta \psi_M = \big\{D_M  - {1 \over 4} H_M \Gamma_{11}
   - {1 \over 4} \, e^{\phi}  m   \,  \Gamma_M  \\[1mm]
 \ - {1 \over 8}  e^{\phi} \big[
  ( \Gamma_{M} F  - 4 F_M )  \Gamma_{11} 
 -{1 \over 12} ( \Gamma_M  G - 12 G_M) \big] \big\} \epsilon 
\\[2mm]  \label{041}
\delta \lambda =  \big\{ {1 \over 2}  \partial \phi 
      - {1 \over 12}  H  \Gamma_{11}
  + e^{\phi}\big[ {5 \over 4} m - {3 \over 8} F  \Gamma_{11}
      - {1 \over 96} G
\big] \big\}  \epsilon 
\ea
\ee
[$\partial \equiv \Gamma^M \partial_M$, $F \equiv
F_{MN} \Gamma^{MN}$].  Apart from the differential forms that we
introduced already, the mass parameter is denoted by $m$ and $\phi$ is
the dilaton.  In type IIA string theory, the Killing spinor $\epsilon$
is Majorana and can be decomposed into two Majorana-Weyl spinors of
opposite chirality. The massless case can be lifted to M-theory and
this Majorana spinor becomes the 11-dimensional Killing spinor.

We are interested in the compactifications to a 4-d spacetime that is
either flat or anti deSitter, i.e. up to warping the 10-d space time
factorizes $M_{10} = X_{1,3} \times Y_6$ and we write the metric as
\be060 
ds^2 = e^{-2V(y)} \, ds_4^{(AdS)} + 
e^{2U(y)} \, h_{mn}(y) \, dy^m dy^n  \ .
\ee
where $h_{mn}$ is the metric on $Y_6$ and the warp factors may depend
only on the coordinates of the internal space. Consistent with this
metric Ansatz is the assumption that the fluxes associated with the
forms $F$ and $H$ have non-zero components only in the internal space
$Y_6$ whereas $G$ may have in addition a Freud-Rubin parameter
$\lambda$:
\be062
\ba{rcl}
F &=& F_{mn} dy^m \wedge dy^n \  , \\[1mm]
H &=& H_{mnp} dy^m \wedge dy^n \wedge dy^p \ , \\[1mm]
G &=& \lambda \, dx^0 \wedge dx^1 \wedge dx^2 \wedge dx^3 
\\
&& + G_{mnpq} \, dy^m \wedge dy^n \wedge dy^p \wedge dy^q \ .
\ea
\ee
Note,  all forms as well as the warp factor and the dilaton are in
general functions of internal coordinates $y^m$.

Keeping four supercharges unbroken, the 4-d vacuum should allow for a
single Majorana or Weyl (Killing) spinor and the geometry of the 6-d
internal space for a single (Weyl) Killing spinor.  This internal
spinor has to be a singlet of the structure group $G \subseteq$ SU(3)
$\subset$ SO(6): for the case $G=$SU(3), only a single 6-dimensional
Weyl Killing spinor can exist whereas for $G\subset$ SU(3) more
singlet spinors are possible. Actually, a reduced structure group is
equivalent to the existence of vector field(s) on $Y_6$ and puts
severe constraints on the geometry of the internal space. If there are
no fluxes, the Killing spinors are covariantly constant and the
structure group determines the holonomy of the space.  On the other
hand, non-trivial fluxes, which decompose into representations of
the structure group, act as (intrinsic) torsion components and the
holonomy group of $Y_6$ is in general unrestricted. 

We decompose the 10-d spinor as
\be072
\epsilon = \theta \otimes \eta + \theta^\star \otimes \eta^\star
\ee
where $\theta$ is the 4-d spinor and the 6-d spinor reads
\be138
\eta = {1 \over \sqrt 2}  e^{\alpha + i \beta}
\, ({\mathbb I} - \gamma^7) \, \eta_0 
\ee
with $\alpha$ and $\beta$ as real functions and $\eta_0$ being a
constant spinor. Since we want to keep this spinor as an SU(3)
singlet, it has to obey the projector conditions
\be230
\ba{rcl}
(\gamma_m  - i J_{mn} \gamma^n ) \, \eta &=& 0 \\[1mm]
(\gamma_{mn} +  i\, J_{mn}) \, \eta &=& {i \over 2} \,
e^{2i\beta} \Omega_{mnp} \gamma^p  \, \eta^\star \ ,
\label{231} \\[1mm]
(\gamma_{mnp} + 3 i J_{[mn} \gamma_{p]} ) \, \eta &=& i \, e^{2i\beta}
\Omega_{mnp} \eta^\star
\ea
\ee
with the almost complex structure and  holomorphic 3-form defined by
\be150
\eta \, \gamma_{mn} \eta^\star = i \, e^{2 \alpha}  J_{mn}
\  , \ 
\eta \gamma_{mnp} \eta = i \, e^{2 (\alpha + i\beta) } \Omega_{mnp}
\ee
so that $e^{2\alpha} = \eta^\star \eta$. The phase $\beta$ comes
always together with $\Omega$ and to keep the notation simple, we will
drop this phase and will only comment on it where necessary.  Note,
these are the only differential forms that can be constructed from a
single chiral spinor and for non-zero fluxes they are in general {\em
not} covariantly constant nor closed and this failure is related to
non-vanishing intrinsic torsion components. Following the literature,
see \cite{630,340,Gauntlettetal,Louisetal,LustetalII}, one introduces
five classes ${\cal W}^i$ by
\be524
\ba{rcl}
dJ &=& {3 i \over 4}\,  ( {\cal W}_1 \bar \Omega -
 \bar{\cal W}_1 \Omega ) +  {\cal W}_3 + J \wedge  {\cal W}_4 \ ,\\
d\Omega &=&   {\cal W}_1 J \wedge J + J \wedge  {\cal W}_2
+ \Omega \wedge  {\cal W}_5
\ea
\ee
with the constraints: $J \wedge J \wedge {\cal W}_2 =J \wedge {\cal
W}_3 = \Omega \wedge {\cal W}_3=0 $.  Depending on which torsion
components are non-zero, one can classify the geometry of the internal
space. E.g., if only ${\cal W}_1 \neq 0$ the space is called nearly
K\"ahler, for ${\cal W}_2 \neq 0$ almost K\"ahler, the space is
complex if ${\cal W}_1 ={\cal W}_2= 0$ and it is K\"ahler if only
${\cal W}_5 \neq 0$.  Although it is not possible to introduce complex
coordinates globally, one can nevertheless employ the holomorphic
projector: ${1 \over 2} (\1 \pm i J)$ to distinguish between
holomorphic and anti-holomorphic indices locally. This is useful to
decide whether flux components can cancel or not.

The 4-d spinor can be Weyl or Majorana. A detailed analysis shows for
the Weyl case \cite{333}, that the mass and all RR-fields have to
vanish and only the $H$-flux can be non-zero; see also
\cite{LustetalII,670}. So, we shall consider a 4-d Majorana spinor
implying that the 10-d spinor is not chiral, which is generic for an
type IIA vacuum. Hence,
\be211
e^{i \beta } \theta \equiv \hat \theta = \hat \theta^\star  \ ,
\ee
where the phase $\beta$ describes the mixing of the two chiralities of
the 10-d spinor (\ref{072}); for $\beta = 0$ both chiralities are on
equal footing\footnote{In order to solve the Killing spinor equations
we had to identified $\beta$ with the phase factor for the holomorphic
3-form in (\ref{150}).}.  Since we allow for an external
AdS-space, this spinor obeys
\be399
\nabla_\mu  \, \theta = 
\hat \gamma_\mu  (W_1 + i \hat \gamma^5 W_2) \, \theta
\ee
where the 4-d $\gamma$-matrices are hatted. Obviously, upon
dimensional reduction $W_{1/2}$ (not to be confused with the torsion
classes ${\cal W}_i$) will fix the real and imaginary part of the
superpotential and its absolute value acts as cosmological constant
yielding an anti deSitter vacuum.

Now, we have to separate all terms containing $\theta$ from the terms
proportional to $\hat\gamma^5 \theta$.  The gravitino variation
(\ref{040}) is spited into an external and internal part and
together with dilatino variation, each supersymmetry variation yields
a constraint equation on the fluxes and one differential equation.
Collecting all terms of the same 6-d chirality, the constraint
equations can be written as
\be229
\ba{rcl}
{1 \over 4} e^{\phi}
       (m - {1 \over 24} e^{-4U} G)\eta &=&  e^V\, W_1 \eta    \\ 
  e^{\phi} (15 m - {1 \over 8}
    e^{-4U} G) \eta  &=& e^{-3U}\, H \, \eta^\star  \\  
  4\,  e^{U+V} W_1 \gamma_m \eta+  
	  {1 \over 2} e^{\phi- 3U} G_m  \eta &=&  
   e^{-2U} \, H_m \, \, \eta^\star  \label{228}
\ea
\ee
whereas for the differential equations we derive
\be222
\ba{l}
   e^{-U} \partial V \eta = {1 \over 4} (e^\phi [e^{-2U} F + {i \over 6}
e^{4V} \lambda] + 2 i  W_2 e^{V} ) \eta^\star   \\[1mm]  \label{223}
  e^{-U} \partial \phi \, \eta = 
     - {3\over 4} 
e^\phi (e^{-2U} F + {i \over 36}  e^{4V} \, \lambda ) \, \eta^\star \\  
e^{-U}   \hat \nabla_m \eta = {1 \over 2}  e^{\phi} (e^{-2U} F_m + 
{i \over 16} e^{4V} 
\, \lambda\, \gamma_m\,) \eta^\star  \\[1mm]
\qquad \qquad \qquad  + i e^{V} \,  W_2 \,   \gamma_m \eta^\star  
\ea
\ee
where: $\hat\nabla_m \equiv \nabla_m + {1 \over 2}
\gamma_m{}^n\partial_n (U +V) + {1 \over 2} \partial_m U$.  Recall,
the indices of $G$, $F$ etc., are contracted with the
$\gamma$-matrices, but using the relations (\ref{230}) it follows from
(\ref{228}) that the only non-zero components of the fluxes are given
by
\be774
G = G_0 \, J \wedge J \ , \ H = H_0 \, {\rm Im} \Omega
\ee
which are the singlet components  under an SU(3) decomposition.
In addition, we infer
\be671
W_1 = 0 \ , \ G_0 = -8 \, e^{4U}m \ , \ H_0 = 12 \, e^{3U+\phi} m 
\ee  
Eq.\  (\ref{222}) yield
\be882
F_0 = - {1 \over 36} \lambda e^{2U + 4V} 
\ , \ W_2 = - {5 \over 72} \lambda\, e^{\phi + 3V}
\ee
and moreover
\be661
\ba{rcl}
2 \nabla_m \hat \eta &=&   -  \gamma_m{}^n  \partial_n (U+V) \, \eta 
+ e^{\phi - U} F_{mn} \gamma^n\, \hat \eta^\star \\[1mm] \label{663}
e^U \partial_m V &=& -{i \over 8} e^\phi \, \Omega_{mpq} F^{pq}
\ea
\ee
with $\hat \eta = e^{U \over 2} \eta$.  Next, the 10-d equations of
motion for $G$ and $H$ imply
\be227
d(^\star G) \sim G \wedge H \ , \quad 
d(e^{-2\phi} \, {^\star H }) \sim G\wedge G \ .
\ee
The rhs is non-zero only if the Freud-Rubin parameter is non-zero and
when projected onto the internal components we find from (\ref{774}):
$^\star G \sim J$, $^\star H \sim {\rm Re} \Omega$. Going back to our
torsion classes as introduced in (\ref{524}), the equations of motion
can be solved only if
\[
{\cal W}_2 = {\cal W}_3 = 0 \ .
\]
Next, by a proper conformal rescaling of the internal manifold, one
can eliminate ${\cal W}_4$ (see \cite{340}), which amounts to a proper
choice for the warp factor $U$. Using the differential equations
(\ref{661}) to calculate $dJ$ and $d\Omega$ we find that this is
consistent if: $U=V$.  This requirement would also be consistent with
the Bianchi identity $dG = 6 F \wedge G$, but does not solve the
equations of motion (\ref{227}) (which would require $U=-2V$). The
only solution that we found requires
\be557
d\phi = dV = dU = 0 \ , \quad F= F_0 J
\ee
with constant $G_0$, $H_0$ given by (\ref{671}) and $F_0$ by
(\ref{882}).  Note, $U$ and $V$ can be eliminated by an rescaling of
the external and internal coordinates combined by a rescaling of the
cosmological constant $W_2 \sim \lambda$.  Hence, setting $U=V=0$ from
the very beginning, we find that the dilaton is fixed by the ratio of
the (quantized) fluxes
\be837
e^\phi = -{2 \, H_0 \over 3 \, G_0} \ .
\ee
The differential equation for the spinor becomes finally
\be823
\nabla_m \eta = - {i \over 72}  e^\phi \lambda \gamma_m \eta^\star
\ee
which yields (with $\beta = \alpha = 0$)
\[
{\cal W}_1 = {\lambda \over 54} e^\phi \ , \quad 
{\cal W}_2 ={\cal W}_3 ={\cal W}_4 ={\cal W}_5 =0 \ .
\]
This identifies the internal space as a nearly K\"ahler manifold, which
is Einstein but neither complex nor K\"ahler! In fact, $dJ = -{\lambda
\over 36} e^\phi {\rm Im} \Omega$, $d\, {\rm Re} \Omega \sim J \wedge
J$ which ensures $dF = m H$, $dH = 0$ and $d\, {^\star H} \sim G
\wedge G$. This also fixes the Freud-Rubin parameter in terms of the
mass: 
\[
{\lambda \over 72} = \sqrt{3} m \ .
\]
In the limit of vanishing mass, our solution becomes trivial, i.e.\
all fluxes vanish and the internal space becomes Calabi-Yau. There is
however no direct limit to massless configurations related to
intersecting D6-branes, which have ${\cal W}_1 ={\cal W}_3 =0$, but
${\cal W}_2, {\cal W}_4, {\cal W}_5\neq 0$ \cite{351}. Note, the
external space is anti deSitter with the cosmological constant given
by the Freud-Rubin parameter, which in turn is related to the
mass. The differential equation of the spinor can be solved by a
constant spinor if one imposes first order differential equations on
the Vielbeine $e^n$
\be525
\omega^{pq} J_{pq} = 0 \ , \quad 
\omega^{pq} \Omega_{pq}{}^n = - {\lambda \over 9} e^\phi e^n 
\ee
where $\omega^{pq} \equiv \omega^{pq}_m dy^m$ are the spin-connection
1-forms. Therefore, 6-dimensional nearly K\"ahler spaces can be seen
as a weak SU(3)-holonomy space, which as Calabi-Yau spaces have, e.g.,
a vanishing first Chern class.  Their close relationship to special
holonomy spaces comes also due to the fact that the cone over nearly
K\"ahler 6-manifolds become a $G_2$-holonomy spaces \cite{690},
defined by a covariantly constant spinor. This can be verified by
multiplying (\ref{525}) with $J$ from the right and identifying the
rhs as the spin connection $\omega^{7n}$.  Note, the spin connection
1-form of $G_2$ holonomy spaces satisfy $\omega^{MN} \varphi_{MNP} =
0$, where $\varphi_{MNP}$ is the $G_2$-invariant 3-form. It is hence
straightforward to construct nearly K\"ahler spaces starting from
$G_2$ holonomy spaces and the almost K\"ahler form, that defines our
vacuum completely, is then given by $J_{mn} = \varphi_{mn7}$.  Let us
end with a discussion of some coset examples; for more details we
refer to \cite{710, 760}.

$(i)\ {G_2 \over SU(3)} \simeq S_6$ \  This is a standard
example of a nearly K\"ahler space, where the cone becomes the flat
7-d space. Note, one can express the 6-sphere also by the coset
${SO(7)/SO(6)}$ which however breaks supersymmetry.

$(ii)\ {Sp(2) \over Sp(1) \times U(1)} \simeq {S_7 \over U(1)} \simeq
{\mathbb CP_3}$ \ The corresponding $G_2$-holonomy space is an
${\mathbb R_3}$ bundle over $S_4$ and hence it is the $SO(5)$
invariant metric of ${\mathbb CP_3}$ appearing here and not the
$SU(4)$-invariant, which is K\"ahler (instead of nearly
K\"ahler) and hence would break supersymmetry.

$(iii)\ {SU(3) \over U(1) \times U(1)}$ \ The cone over this space
gives the $G_2$-holonomy space related to an ${\mathbb R_3}$ bundle
over ${\mathbb CP_2}$ and therefore the 6-dimensional metric is
$SU(3)$-invariant. This space is isomorphic to the flag manifold,
which again allows for another metric which is K\"ahler and 
would break supersymmetry.

$(iv)\ {SU(2)^3 \over SU(2)} \simeq S_3 \times S_3$ \ There are
different possibilities of modding out the $SU(2)$ and the nearly
K\"ahler space appearing in our context is obtained by a diagonal
embedding yielding as $G_2$ manifold an ${\mathbb R_4}$ bundle over
$S_3$.

Having identified the 7-dimensional space with $G_2$-holonomy it is
straightforward to obtain the metric and the almost complex structure
$J$ of the nearly K\"ahler 6-manifold; see \cite{720,730}. Actually
there is also a whole class of known non-homogeneous (singular)
examples, which are obtained from $G_2$ manifold given by an ${\mathbb
R_3}$ bundle over any 4-d selfdual Einstein space, where the nearly
K\"ahler space becomes an $S_2$ bundle over the 4-d Einstein space,
which is also known as the twistor space; $(i)$ and $(ii)$ are just
the simplest (regular) examples, see also \cite{750, 740}.  Having
related the nearly K\"ahler space to $G_2$-manifolds, it is tempting
to identify the $7^{th}$ direction with the radial direction of the
AdS space so that the 10-d metric can thus be interpreted as the
near-horizon geometry of a massive D2-brane. The transversal space is
a (conical) $G_2$-manifold accommodating the NS-NS and RR-fluxes
(\ref{774}) and it breaks 1/8 of supersymmetry.

It is obvious to ask also for brane sources consistent with this
background. This question deserves of course a detailed analysis, but
the topology of the spaces suggest already a number of interesting
candidates. E.g.\ 4- and 6-branes on ${\mathbb CP_3}$ might be wrapped
on 2- and 4-cycles yielding a domain wall in the external space.  But
more interesting {from} the field theory perspective are branes that
extend along the whole external space time, which would be the case if
one can wrap 6-branes around each $S_3$ of the coset ${SU(2)^3 \over
SU(2)}$.  In either case the mass parameter can be related to 8-branes
wrapping the 6-manifold and extending in two external directions
yielding the $AdS_4$ geometry.


\begin{acknowledgments}
Research is supported in part by DOE grant DOE-EY-76-02-3071 (M.C.),
NSF grant INT02-03585 (M.C.) and Fay R. and Eugene L. Langberg endowed
Chair (M.C.) and a Heisenberg Fellowship (K.B.).
\end{acknowledgments}



\begin{thebibliography}{24}
\expandafter\ifx\csname natexlab\endcsname\relax\def\natexlab#1{#1}\fi
\expandafter\ifx\csname bibnamefont\endcsname\relax
  \def\bibnamefont#1{#1}\fi
\expandafter\ifx\csname bibfnamefont\endcsname\relax
  \def\bibfnamefont#1{#1}\fi
\expandafter\ifx\csname citenamefont\endcsname\relax
  \def\citenamefont#1{#1}\fi
\expandafter\ifx\csname url\endcsname\relax
  \def\url#1{\texttt{#1}}\fi
\expandafter\ifx\csname urlprefix\endcsname\relax\def\urlprefix{URL }\fi
\providecommand{\bibinfo}[2]{#2}
\providecommand{\eprint}[2][]{\url{#2}}

\bibitem[{\citenamefont{Blumenhagen et~al.}(2001)\citenamefont{Blumenhagen,
  {K\"ors}, and {L\"ust}}}]{Blumetal}
\bibinfo{author}{\bibfnamefont{R.}~\bibnamefont{Blumenhagen}},
  \bibinfo{author}{\bibfnamefont{B.}~\bibnamefont{{K\"ors}}}, \bibnamefont{and}
  \bibinfo{author}{\bibfnamefont{D.}~\bibnamefont{{L\"ust}}},
  \bibinfo{journal}{JHEP} \textbf{\bibinfo{volume}{02}}, \bibinfo{pages}{030}
  (\bibinfo{year}{2001}), \eprint{hep-th/0012156}.

\bibitem[{\citenamefont{Aldazabal et~al.}(2001)\citenamefont{Aldazabal, Franco,
  Ib\'a\~nez, Rabadan, and Uranga}}]{Ibanezetal}
\bibinfo{author}{\bibfnamefont{G.}~\bibnamefont{Aldazabal}},
  \bibinfo{author}{\bibfnamefont{S.}~\bibnamefont{Franco}},
  \bibinfo{author}{\bibfnamefont{L.~E.} \bibnamefont{Ib\'a\~nez}},
  \bibinfo{author}{\bibfnamefont{R.}~\bibnamefont{Rabadan}}, \bibnamefont{and}
  \bibinfo{author}{\bibfnamefont{A.~M.} \bibnamefont{Uranga}},
  \bibinfo{journal}{JHEP} \textbf{\bibinfo{volume}{02}}, \bibinfo{pages}{047}
  (\bibinfo{year}{2001}), \eprint{hep-ph/0011132}.


\bibitem[{\citenamefont{L{\"u}st}(2004)}]{770}
\bibinfo{author}{\bibfnamefont{D.}~\bibnamefont{L{\"u}st}}
  (\bibinfo{year}{2004}), \eprint{hep-th/0401156}.

\bibitem[{\citenamefont{Cveti{\v c}
  et~al.}(2001{\natexlab{a}})\citenamefont{Cveti{\v c}, Shiu, and
  Uranga}}]{CSUI}
\bibinfo{author}{\bibfnamefont{M.}~\bibnamefont{Cveti{\v c}}},
  \bibinfo{author}{\bibfnamefont{G.}~\bibnamefont{Shiu}}, \bibnamefont{and}
  \bibinfo{author}{\bibfnamefont{A.~M.} \bibnamefont{Uranga}},
  \bibinfo{journal}{Phys. Rev. Lett.} \textbf{\bibinfo{volume}{87}},
  \bibinfo{pages}{201801} (\bibinfo{year}{2001}{\natexlab{a}}),
  \eprint{hep-th/0107143}.

\bibitem[{\citenamefont{Cveti{\v c}
  et~al.}(2001{\natexlab{b}})\citenamefont{Cveti{\v c}, Shiu, and
  Uranga}}]{CSUII}
\bibinfo{author}{\bibfnamefont{M.}~\bibnamefont{Cveti{\v c}}},
  \bibinfo{author}{\bibfnamefont{G.}~\bibnamefont{Shiu}}, \bibnamefont{and}
  \bibinfo{author}{\bibfnamefont{A.~M.} \bibnamefont{Uranga}},
  \bibinfo{journal}{Nucl. Phys.} \textbf{\bibinfo{volume}{B615}},
  \bibinfo{pages}{3} (\bibinfo{year}{2001}{\natexlab{b}}),
  \eprint{hep-th/0107166}.

\bibitem[{\citenamefont{Gauntlett et~al.}(2003)\citenamefont{Gauntlett,
  Martelli, and Waldram}}]{650}
\bibinfo{author}{\bibfnamefont{J.~P.} \bibnamefont{Gauntlett}},
  \bibinfo{author}{\bibfnamefont{D.}~\bibnamefont{Martelli}}, \bibnamefont{and}
  \bibinfo{author}{\bibfnamefont{D.}~\bibnamefont{Waldram}}
  (\bibinfo{year}{2003}), \eprint{hep-th/0302158}.

\bibitem[{\citenamefont{Berkooz et~al.}(1996)\citenamefont{Berkooz, Douglas,
  and Leigh}}]{130}
\bibinfo{author}{\bibfnamefont{M.}~\bibnamefont{Berkooz}},
  \bibinfo{author}{\bibfnamefont{M.~R.} \bibnamefont{Douglas}},
  \bibnamefont{and} \bibinfo{author}{\bibfnamefont{R.~G.} \bibnamefont{Leigh}},
  \bibinfo{journal}{Nucl. Phys.} \textbf{\bibinfo{volume}{B480}},
  \bibinfo{pages}{265} (\bibinfo{year}{1996}), \eprint{hep-th/9606139}.

\bibitem[{\citenamefont{Behrndt and Cveti{\v c}}(2004{\natexlab{a}})}]{333}
\bibinfo{author}{\bibfnamefont{K.}~\bibnamefont{Behrndt}} \bibnamefont{and}
  \bibinfo{author}{\bibfnamefont{M.}~\bibnamefont{Cveti{\v c}}},
  \bibinfo{journal}{UPR-1071-T}  (\bibinfo{year}{2004}{\natexlab{a}}).

\bibitem[{\citenamefont{Behrndt and Cveti{\v c}}(2004{\natexlab{b}})}]{100}
\bibinfo{author}{\bibfnamefont{K.}~\bibnamefont{Behrndt}} \bibnamefont{and}
  \bibinfo{author}{\bibfnamefont{M.}~\bibnamefont{Cveti{\v c}}},
  \bibinfo{journal}{Nucl. Phys.} \textbf{\bibinfo{volume}{B676}},
  \bibinfo{pages}{149} (\bibinfo{year}{2004}{\natexlab{b}}),
  \eprint{hep-th/0308045}.

\bibitem[{\citenamefont{Romans}(1986)}]{110p}
\bibinfo{author}{\bibfnamefont{L.~J.} \bibnamefont{Romans}},
  \bibinfo{journal}{Phys. Lett.} \textbf{\bibinfo{volume}{B169}},
  \bibinfo{pages}{374} (\bibinfo{year}{1986}).

\bibitem[{\citenamefont{Grey and Hervella}(1980)}]{630}
\bibinfo{author}{\bibfnamefont{A.}~\bibnamefont{Grey}} \bibnamefont{and}
  \bibinfo{author}{\bibfnamefont{L.}~\bibnamefont{Hervella}},
  \bibinfo{journal}{Ann. Math. Pura Appl.} \textbf{\bibinfo{volume}{123}},
  \bibinfo{pages}{35} (\bibinfo{year}{1980}).

\bibitem[{\citenamefont{Chiossi and Salamon}(2002)}]{340}
\bibinfo{author}{\bibfnamefont{S.}~\bibnamefont{Chiossi}} \bibnamefont{and}
  \bibinfo{author}{\bibfnamefont{S.}~\bibnamefont{Salamon}}
  (\bibinfo{year}{2002}), \eprint{math.dg/0202282}.

\bibitem[{\citenamefont{Gurrieri et~al.}(2003)\citenamefont{Gurrieri, Louis,
  Micu, and Waldram}}]{Louisetal}
\bibinfo{author}{\bibfnamefont{S.}~\bibnamefont{Gurrieri}},
  \bibinfo{author}{\bibfnamefont{J.}~\bibnamefont{Louis}},
  \bibinfo{author}{\bibfnamefont{A.}~\bibnamefont{Micu}}, \bibnamefont{and}
  \bibinfo{author}{\bibfnamefont{D.}~\bibnamefont{Waldram}},
  \bibinfo{journal}{Nucl. Phys.} \textbf{\bibinfo{volume}{B654}},
  \bibinfo{pages}{61} (\bibinfo{year}{2003}), \eprint{hep-th/0211102}.

\bibitem[{\citenamefont{Cardoso et~al.}(2003)}]{LustetalII}
\bibinfo{author}{\bibfnamefont{G.~L.} \bibnamefont{Cardoso}}
  \bibnamefont{et~al.}, \bibinfo{journal}{Nucl. Phys.}
  \textbf{\bibinfo{volume}{B652}}, \bibinfo{pages}{5} (\bibinfo{year}{2003}),
  \eprint{hep-th/0211118}.

\bibitem[{\citenamefont{Gauntlett et~al.}(2002)\citenamefont{Gauntlett,
  Martelli, Pakis, and Waldram}}]{Gauntlettetal}
\bibinfo{author}{\bibfnamefont{J.~P.} \bibnamefont{Gauntlett}},
  \bibinfo{author}{\bibfnamefont{D.}~\bibnamefont{Martelli}},
  \bibinfo{author}{\bibfnamefont{S.}~\bibnamefont{Pakis}}, \bibnamefont{and}
  \bibinfo{author}{\bibfnamefont{D.}~\bibnamefont{Waldram}}
  (\bibinfo{year}{2002}), \eprint{hep-th/0205050}.

\bibitem[{\citenamefont{Becker et~al.}(2003)\citenamefont{Becker, Becker,
  Dasgupta, and Green}}]{670}
\bibinfo{author}{\bibfnamefont{K.}~\bibnamefont{Becker}},
  \bibinfo{author}{\bibfnamefont{M.}~\bibnamefont{Becker}},
  \bibinfo{author}{\bibfnamefont{K.}~\bibnamefont{Dasgupta}}, \bibnamefont{and}
  \bibinfo{author}{\bibfnamefont{P.~S.} \bibnamefont{Green}},
  \bibinfo{journal}{JHEP} \textbf{\bibinfo{volume}{04}}, \bibinfo{pages}{007}
  (\bibinfo{year}{2003}), \eprint{hep-th/0301161}.

\bibitem[{\citenamefont{Kaste et~al.}(2003)\citenamefont{Kaste, Minasian,
  Petrini, and Tomasiello}}]{351}
\bibinfo{author}{\bibfnamefont{P.}~\bibnamefont{Kaste}},
  \bibinfo{author}{\bibfnamefont{R.}~\bibnamefont{Minasian}},
  \bibinfo{author}{\bibfnamefont{M.}~\bibnamefont{Petrini}}, \bibnamefont{and}
  \bibinfo{author}{\bibfnamefont{A.}~\bibnamefont{Tomasiello}}
  (\bibinfo{year}{2003}), \eprint{hep-th/0301063}.

\bibitem[{\citenamefont{B{\"a}r}(1993)}]{690}
\bibinfo{author}{\bibfnamefont{C.}~\bibnamefont{B{\"a}r}},
  \bibinfo{journal}{Comm. Math. Phys.} \textbf{\bibinfo{volume}{154}},
  \bibinfo{pages}{509} (\bibinfo{year}{1993}).

\bibitem[{\citenamefont{Gray}(1970)}]{710}
\bibinfo{author}{\bibfnamefont{A.}~\bibnamefont{Gray}}, \bibinfo{journal}{J.
  Diff. Geom.} \textbf{\bibinfo{volume}{4}}, \bibinfo{pages}{283}
  (\bibinfo{year}{1970}).

\bibitem[{\citenamefont{Atiyah and Witten}(2003)}]{760}
\bibinfo{author}{\bibfnamefont{M.}~\bibnamefont{Atiyah}} \bibnamefont{and}
  \bibinfo{author}{\bibfnamefont{E.}~\bibnamefont{Witten}},
  \bibinfo{journal}{Adv. Theor. Math. Phys.} \textbf{\bibinfo{volume}{6}},
  \bibinfo{pages}{1} (\bibinfo{year}{2003}), \eprint{hep-th/0107177}.

\bibitem[{\citenamefont{Bryant and Salamon}(1989)}]{720}
\bibinfo{author}{\bibfnamefont{R.}~\bibnamefont{Bryant}} \bibnamefont{and}
  \bibinfo{author}{\bibfnamefont{S.}~\bibnamefont{Salamon}},
  \bibinfo{journal}{Duke Math. J.} \textbf{\bibinfo{volume}{58}},
  \bibinfo{pages}{829} (\bibinfo{year}{1989}).

\bibitem[{\citenamefont{Gibbons et~al.}(1990)\citenamefont{Gibbons, Page, and
  Pope}}]{730}
\bibinfo{author}{\bibfnamefont{G.~W.} \bibnamefont{Gibbons}},
  \bibinfo{author}{\bibfnamefont{D.~N.} \bibnamefont{Page}}, \bibnamefont{and}
  \bibinfo{author}{\bibfnamefont{C.~N.} \bibnamefont{Pope}},
  \bibinfo{journal}{Commun. Math. Phys.} \textbf{\bibinfo{volume}{127}},
  \bibinfo{pages}{529} (\bibinfo{year}{1990}).

\bibitem[{\citenamefont{Cveti{\v c} et~al.}(2003)\citenamefont{Cveti{\v c},
  Gibbons, Lu, and Pope}}]{750}
\bibinfo{author}{\bibfnamefont{M.}~\bibnamefont{Cveti{\v c}}},
  \bibinfo{author}{\bibfnamefont{G.~W.} \bibnamefont{Gibbons}},
  \bibinfo{author}{\bibfnamefont{H.}~\bibnamefont{Lu}}, \bibnamefont{and}
  \bibinfo{author}{\bibfnamefont{C.~N.} \bibnamefont{Pope}},
  \bibinfo{journal}{Class. Quant. Grav.} \textbf{\bibinfo{volume}{20}},
  \bibinfo{pages}{4239} (\bibinfo{year}{2003}), \eprint{hep-th/0206151}.

\bibitem[{\citenamefont{Behrndt et~al.}(2002)\citenamefont{Behrndt, Dall'Agata,
  L{\"u}st, and Mahapatra}}]{740}
\bibinfo{author}{\bibfnamefont{K.}~\bibnamefont{Behrndt}},
  \bibinfo{author}{\bibfnamefont{G.}~\bibnamefont{Dall'Agata}},
  \bibinfo{author}{\bibfnamefont{D.}~\bibnamefont{L{\"u}st}}, \bibnamefont{and}
  \bibinfo{author}{\bibfnamefont{S.}~\bibnamefont{Mahapatra}},
  \bibinfo{journal}{JHEP} \textbf{\bibinfo{volume}{08}}, \bibinfo{pages}{027}
  (\bibinfo{year}{2002}), \eprint{hep-th/0207117}.

\end{thebibliography}


\end{document}